# Hidden and Uncontrolled – On the Emergence of Network Steganographic Threats

Steffen Wendzel[1], Wojciech Mazurczyk[2],
Luca Caviglione[3], Michael Meier[1]

[1] Cyber Defense Research Group, Fraunhofer FKIE, Bonn, Germany
{steffen.wendzel, michael.meier}@fkie.fraunhofer.de

[2] Institute of Telecomm., Warsaw University of Technology, Warsaw, Poland
wmazurcz@elka.pw.edu.pl

[3] Institute of Intelligent Systems for Automation (ISSIA),
National Research Council of Italy (CNR), Genoa, Italy
luca.caviglione@ge.issia.cnr.it

## Abstract

Network steganography is the art of hiding secret information within innocent network transmissions. Recent findings indicate that novel malware is increasingly using network steganography. Similarly, other malicious activities can profit from network steganography, such as data leakage or the exchange of pedophile data. This paper provides an introduction to network steganography and highlights its potential application for harmful purposes. We discuss the issues related to countering network steganography in practice and provide an outlook on further research directions and problems.

## 1 Introduction

Steganography is known to be a technology used since thousands of years; its purpose is to embed a secret message into an innocent looking carrier. Digital media steganography embeds secret data into digital structures, including digital videos, digital audio files, and digital image files. Within recent years, a novel part of steganography arose, namely *network steganography*. Network steganography transfers secret data over a network by hiding secret information into legitimately appearing transmissions [LuWS10].

In comparison to cryptography, steganography aims at hiding the existence of a secret message while cryptography aims on hiding the content of a message. Both technologies cryptography and steganography are orthogonal and can be combined, i.e. a secret message can be encrypted and afterwards it can be hidden using steganography.

Steganography is applied when the detection of the plain *existence* of a secret message can cause harm to the transmitters or another party. For instance, if illicit information is transmitted by journalists over a part of the Internet facing censorship within a dictatorship, the application of network steganography can protect the journalist from being caught while only encrypting the transmission raises attention. As network steganography is a dual-use good, it can also be applied for malicious purposes, including the hidden transmission of command and control (C&C) information for botnets or for the hidden exchange of pedophile content.

In this paper, we first provide a background on network steganography and afterwards highlight recent developments in the field. We show network steganography's potential for malicious activities and discuss the application of countermeasures in the context of these novel approaches.

The remainder of this paper is structured as follows. Section 2 provides fundamentals while Section 3 covers recent developments within the theme. Section 4 provides an outlook on the possible future of network steganography for malicious purposes and Section 5 deals with the question of how to counter network steganography in practice. Section 6 concludes our discussion.

## 2  Fundamentals

Before covering the state of the art in network steganography, this section provides a brief introduction into the theme and highlights selected use-cases for the technology.

### 2.1  Basic Concepts

The term network steganography was coined and first introduced by Szczypiorski in 2003 [Szczy03] and it embraces all the methods that embed secret data within innocent looking network transmissions. Therefore, it is necessary to define a way in which sender and receiver exchange the secret data, i.e. both communicators must agree on a signaling technique in advance and create a so-called *covert channel* over which the hidden data is transmitted. The term covert channel was initially introduced in operating system research by Lampson as a channel not intended for a communication [Lamp73]; later the term was defined as policy-breaking communication channel by the U.S. Department of Defense [Depa85].

Hundreds of techniques can be applied to create a covert channel; some of the most common techniques are to place secret data into unused fields of network protocol headers or to change the size of network packets. Also it is common to manipulate inter packet gaps (IPG), i.e. the time between network packets, and the order of network packets. Moreover, the order of header elements (e.g. HTTP plaintext header lines) can be altered or artificial errors can be introduced into a transmission. A comprehensive analysis and categorization of more than hundred known techniques can be found in [WZFH14].

### 2.2  The Malicious Applications of Network Steganography

The use-cases for network steganography and steganography in general are manifold but as indicated in recent research are mainly linked to illicit purposes [ZiMS14]. A well-known example is the exchange of child pornography using steganography as part of *Operation Twins* in 2002 and the exfiltration of data from the U.S. Department of Justice in 2008 where sensitive financial data was stolen. Other cases are the leakage of classified information from the U.S. to Russia by a Russian spy ring in 2010 and the application of steganography in the *Duqu* and *Alureon* malware [ZiMS14].

While most of these examples were based on traditional digital media steganography (especially digital image steganography) network steganography can be found "in the wild" as well. The rise of network steganography is thereby supported by two factors which make it superior in comparison to digital media steganography [ZiMS14]. Firstly, the space for data to be transferred is not limited in the same way as with a given media object (e.g. an image file) as new data can be transmitted on demand while the capacity of a media object is limited. Secondly, network transmissions are harder to analyze in digital forensics as only parts of the data are stored – if the transmitted data is stored at all.

In particular, network steganographic techniques are available since 1987 in research when first network covert channel techniques were presented [Girl87] and since the mid-90s in the hacking community that brought up tools like *LOKI2* [Daem97] or *Ping Tunnel* [Stod09]. The tools provided by the hacking community embed hidden data in a simple way into the payload of ICMP echo messages or the payload of UDP messages – just to mention a few. The tools are available for free on-line, i.e. accessible for everyone. It was only a matter of time till malware started to apply network steganography to conceal its transmissions. For instance, a recent Linux malware called *Linux.Fokirtor* hides data within SSH transmissions [Prin13].

# 3   Recent Developments in Network Steganography

The recent years of academic research led to even more sophisticated network steganographic techniques. These techniques do not only improve the *stealthiness* (or *covertness*) of the created covert channels but also extensively enhance the capabilities of the steganographic communication [WenK14].

## 3.1   Shift in the Hidden Data Carrier Selection

The predecessors of current sophisticated network steganography methods were utilizing mainly different fields of TCP/IP stack protocols' headers as a hidden data carrier and they focused on the embedding of secret data in the unused or reserved fields. However, recently we have experienced the change in the hidden data carrier selection. Now, the most favorable carrier is the one that is very commonly used so it is hard to spot a single steganographic communication along a vast volume of similar network traffic ("a needle in a hay stack"). Moreover, steganographers tend to exploit higher layers applications and services and the methods are crafted in such a way to exploit their characteristic features.

Recent information hiding solutions utilize as hidden data carrier: *(i)* popular Internet services like Skype, BitTorrent or Google search, *(ii)* new networking environments e.g. cloud computing and technologies like wireless networks and *(iii)* novel communication devices like smartphones.

**Popular Internet Services**

Practically every popular Internet service can be exploited by steganographers if it goes with enough volume of traffic that can be altered to produce the covert channel. One such example is IP telephony service which can be considered a fairly recent discovery however until now a decent number of techniques were proposed [Mazu14].

One of the steganographic methods of this kind is *TranSteg* (Transcoding Steganography), [MaSS12] which is based on the general idea of transcoding (lossy compression) of the voice data from a higher bit rate codec, and thus greater voice payload size, to a lower bit rate codec with smaller voice payload size (which should be performed with the least degradation in voice quality possible). In other words, compression of the covert data is utilized to free space for hidden data bits.

Other methods target even more currently popular peer-to-peer (p2p) services like Skype and BitTorrent. For the first solution a steganographic method named *SkyDe* (Skype Hide) has been proposed by Mazurczyk *et al.* [MaKS2013]. As a hidden data carrier encrypted Skype voice packets are used. By taking advantage of the high correlation between speech activity and packet size, packets without voice signal can be identified and used to carry secret data by replacing the encrypted silence with secret data bits. In the second case the method named *StegTorrent* has been introduced for the BitTorrent application [KoMS13]. It benefits from the fact that, in BitTorrent there are usually many-to-one transmissions, and that for one of its specific protocols – µTP – the header provides means for packets' numbering and retrieving their original sequence. Thus this characteristic feature allows to mark each packet leaving the transmitter regardless of the connection it was sent from so the synchronization is not required. StegTorrent functions based on intentional BitTorrent data packet resorting at the receiver to achieve the desired packet sequences.

It was also proved that information hiding is possible also by simply performing a series of innocent-looking Google searches. The *StegSuggest* steganographic method [BiMS11] targets the feature *Google Suggest* which lists the 10 most popular search phrases given a string of letters a user has entered in Google's search box. The hidden data transmitter intercepts the traffic exchanged between Google's servers and the hidden data receiver's browser when some Google search is initiated. Then, the packets traveling from Google to the hidden data receiver are intercepted and modified by adding a unique word to the end of each of the 10 phrases Google suggests. The choice of phrases is made from a list of 4096 common English words, so the new phrases do not look too suspicious. Then the receiver extracts each added word and converts it into a 10-bit sequence using a previously shared lookup table.

**New Networking Environments and Technologies**

The cloud computing environment can be also exploited to enable covert communication. Ristenpart *et al.* [RTSS09] showed that it is possible to perform cross-virtual machine information leakage by introducing a wide range of techniques for obtaining confidential data by probing the values of shared-cache load, CPU load, keystroke activity, and similar.

Covert communication possibilities have been also recently considered for Content-Centric Networks (CCNs) which are envisaged to replace current IP-based Internet in the future. Ambrosini *et al.* [ACGT14] inspected the potential vulnerabilities that can be exploited for information hiding purposes. They focus on so called *ephemeral* covert communication where the communicating parties are not exchanging any packets directly instead they intentionally influence the content of the router's cache and the delay in its responses to embed and extract hidden data bits. Hidden messages are then present in the cache only for the limited time period and later they are automatically deleted from the network without any additional, required actions.

Information hiding techniques also target wireless networks, are considered a very popular and dynamically evolving network steganography subfield where different standards can be subject to covert communication. For example, for WLANs, Szczypiorski and Mazurczyk have introduced a method called *WiPad* (Wireless Padding) [SzcW11]. The technique is based on the insertion of hidden data into the padding of frames at the physical layer of WLANs. A similar concept was also utilized e.g. for LTE (Long Term Evolution) [GraS14].

Other promising future-network protocols, like, for example, the SCTP (Stream Control Transmission Protocol) which is a candidate for new transport layer protocol, and might replace TCP (Transmission Control Protocol) and UDP (User Datagram Protocol) protocols, turned out to be also prone to steganography. A detailed analysis in [FrMS12a] revealed steganographic vulnerabilities that could be utilized for information hiding with a focus on new features and characteristic of SCTP, such as multi-homing and multi-streaming.

**Novel Communication Devices**

The recent wave of smartphones is equipped with hardware capabilities previously only available in standard desktops or laptops. They now integrate into a unique tool functionalities historically offered by different devices, for instance the camera or the Global Positioning System (GPS). Besides, advancements in wireless connectivity allow connecting to the Internet by using different standards, such as the IEEE 802.11, 3G/4G or the LTE. Therefore, smartphones are now used to exchange and store high volumes of personal data, interact with *Online Social Networks* (OSNs) and in the daily working practice. The latter is also accounting for new paradigms, for instance the *Bring Your Own Device* (BYOD). To handle such complexity, modern smartphones are equipped with an OS having a degree of sophistication comparable to desktop counterparts. In some cases, mobile OSs are ports of desktop-class versions, e.g., Android runs a Linux kernel. However, the vocation of storing sensitive data by-design requires proper additional layers to ensure security and confidentiality of data. While different techniques exist, the most popular idea is to rely on sandboxes, i.e., execution environments forcing a process (application) to not access sensitive data or portion of the hardware.

In this perspective, steganographic possibilities are dramatically multiplied in smartphones for the following reasons: *(i)* the multimedia capability enables to create and use a wide variety of carriers, such as, audio, video, pictures, or Quick-Response (QR) Codes; *(ii)* the availability of a full featured TCP/IP stack, as well as the possibility of interact with desktop-class services lead to a complete reutilization of all the network methods already available for standard computing devices or appliances; *(iii)* the richness of the adopted OS permit to develop sophisticated applications, thus making covert channels based on VoIP and p2p exploitable. However, the most active trend in research and development of steganography methods for smartphones consciously neglect network methods. Firstly, the security layers used in mobile OSs turn out to be barely adequate. Secondly, many manufacturers enforce users to install software only from verified sources (see, e.g., iOS which only permit software provided by the Apple AppStore) that deeply inspect the software before the publishing stage, thus early detecting malicious threats. Summarizing, the real challenge for smartphone steganography is not only "*using network to exfiltrate data*", but "*acquire the data to be exfiltrated*" and "*gain access to the network*" [MazC14].

Therefore, network steganography in smartphones is almost always jointly used with a *local covert channel*. While, the former can be implemented by directly borrowing methods from desktops, the latter is the actual technological enabler for data exfiltration. The paradigmatic use case of a local covert channel, considers two processes, P1 and P2, each one running in a sandbox S1 and S2, respectively. Specifically, S1 and S2 have been configured to prevent data leaking. For instance, if P1 can access the address book, S1 will impede to use network interfaces. Let us assume then P1 is a malware wanting to exfiltrate the address book, while P2 is companion application (often named a colluding application in many research works) looking innocuous. This security mechanism is typically bypassed as follows: P1 leaks data to P2 via covert channel, and P2, after receiving the data, will use network steganography to contact a remote facility in a stealthy manner.

The first work implementing such a mechanism is a malware called *SoundComber* [SZIK11], which is able to capture personal user data, like the digits entered on the keypad of the device during a phone call. It can use up to four different methods to enable the two colluding processes to communicate: 1) by embedding secrets in well-defined patterns of toggling the vibration; 2) by hiding data within changes in the volume level of the ringtone; 3) by locking/unlocking the screen; 4) by concurrently competing in a coordinated flavor on a file lock. Marforio *et al.* [MaFC12] extended this idea by implementing seven additional local covert channels. While some ideas partially overlap with previous ones, their work also exploits a widely used feature of the Android OS: intents. Intents are distributed notifications used to implement basic forms of Inter Process Communication (IPC). Thus, two colluding processes can communicate by exchanging patterns of notifications, e.g., the calendar has changed or the language has been modified. We point out that, while intents are limited to Android, some forms of notifications are also available in iOS, thus making the method applicable in other platforms. Lastly, Lalande and Wendzel [LalS13] offered a variation of the aforementioned mechanisms by using a mix of operations performed on the task list, the screen state, and by manipulating the process priorities to achieve higher undetectability.

## 3.2   Control Protocols and Adaptive Techniques

The 2004 released tool Ping Tunnel [Stod09] is not only capable of transferring hidden data using ICMP echo request and reply but does also place a control protocol header into its payload. Such a control protocol usually implements reliability into the channel, i.e. allowing re-sending lost packets and moreover allowing re-ordering hidden data at the receiver side.

Control protocols for network steganography were discussed in various academic publications in which their feature-set was enhanced.

A significant step in this regard was the automatic discovery of techniques that can be used between two or more peers to exchange secret data. Yarochkin *et al.* introduced a *network environment learning* (NEL) phase in which peers probe the hiding of data within different network protocols using a set of methods and filter out blocked and non-routed network protocols [YDL+08]. This also allows bypassing administrative set-ups and changes in the configuration. For instance, if a network steganographic communication got detected and the administrator blocks the utilized communication it can automatically switch to another hiding method or another network protocol to establish a network steganographic communication nevertheless [YDL+08, WenK14, JaWS13].

The next step for network steganography was to build hidden overlay networks capable of realizing a routing process. The first approach – based on the random-walk algorithm – was already presented in 2007 by Szczypiorski *et al.* in [SzIW07]. Later, a stealth-optimized dynamic routing based on the optimized link-state routing (OLSR) using size-optimized control protocols was published [BaWK12].

## 3.3   Optimization Means

While control protocols for network steganography increased the capabilities of hidden communication, a following step was to optimize not only undetectability of the transmission of the user's secret data bits but the control protocol's stealthiness, too. This concept was initially discussed in [WenK11], where the optimal size of a control protocol header for simultaneously used carrier protocols was discussed [WenK11]. Moreover, two additional protocol engineering approaches for control protocols are available today.

The *first* approach presented in [WenK12] utilizes formal grammar to adjust the embedded control protocol to the utilized network protocol in which it is hidden. Therefore the specification of the hidden control protocol must be designed in a way that it matches the specification of the utilized protocol in a way that no anomaly is caused by the hidden control protocol's operation.

The less secret data is transferred by a steganographic method, the less attention will be raised. Therefore, the *second* approach for control protocol engineering is to minimize the size of the control protocol to as few bits as possible, what can also be achieved by separating the components of the protocol [BaWK12].

For optimizing the communication process between peers, approaches exist to minimize the caused overhead of a covert communication as well as to minimize the packet count for a given amount of bytes to be transmitted over simultaneously used carriers [WenK11].

An additional view on the improvement of the undetectability for covert communication was presented by Frączek *et al.* [FrMS11]. Besides classifying the techniques to improve the stealthiness of a hidden communication, the authors introduced *multi-level steganography* (MLS) in which at least two steganographic methods are utilized simultaneously in such a way that one method's network traffic serves as a carrier for the second method (previous approaches did only combine unnested parallel connections [WenK11]) Such a relationship has been proven to have several potential benefits: *(i)* the lower-level method can carry a cryptographic key that deciphers the steganogram carried by the upper-level method; *(ii)* it can be used to provide the steganogram with integrity, and *(iii)* it may be assigned as a signaling channel of the control protocol [FrMS12b].

The most recent idea is to take the distortion of the carrier that is used for transferring hidden data into account. The so-called *network steganographic cost* measures this distortion and can be compared to measures such as the Peak Signal-to-Noise Ratio (PSNR) or Mean Opinion Score (MOS) known from digital media steganography [MWIS14]. By optimizing the cost of a network steganographic transmission, the stealthiness of the communication can also be increased.

## 4 Outlook

While the previous section covered novel approaches for network steganography, including its sub-domain network covert channels, we will now provide an outlook on the possible future applications of network steganography for malicious purposes.

Firstly, we expect that the application of the mentioned techniques leads to **more sophisticated malware**. This means that malware will be stealthier and thus harder to detect than today forming a new type of *Advanced Persistent Threat* (APT). Deploying dynamic overlay routing and adaptive network covert channel techniques, malware will moreover be harder to prevent on the network level. A host-based detection of malware is thus important when a network-level detection and prevention is not feasible in these situations. Today's command and control (C&C) channels in botnets already possess a comprehensive feature set and adapting these features to the context of network steganography is only one step in the malware evolution.

Secondly, we observe the **increasing stealthiness of malware communications on smart phones**. While there is not a major effort in developing novel network covert channels especially crafted for smartphones, recent trends take advantage on the device offloading features, especially those using the cloud. In fact, to bypass some storage or battery limitation of devices, some operations are delegated to a remote server farm. In this perspective, new applications producing traffic potentially exploitable for network steganography are becoming available. We mention among the others, voice-based services like Google Now and Siri or cloud storage platforms like Google Drive and Dropbox. Even if many frameworks use protocols/techniques already used for network steganography (e.g., HTTP), the huge volumes and the degree of sophistication of many services will represent a great challenge, especially in terms of being able to detect the covert communication or to provide effective countermeasures.

Thirdly, we predict the **emergence of network steganography to new domains**, especially when combined with existing malware. One example in this regard is the potential to form novel botnets consisting of smart buildings instead of computers (so called *smart building botnets*), allowing the remote mass-surveillance and remote control of smart cities [WZMS14]. Network steganography can increase the stealthiness of mass surveillance in such situations, especially when the number of bots in a smart building botnet is high.

Fourth, network steganography could **increase the stealthiness of illegal data exchange**, including communications we find within darknets already today, such as the exchange of child pornographic material or hitman hiring.

Fifth, it can be expected that network steganography can highly **influence to industrial espionage when it comes to data leakage**. While today's data leakage is performed quite plainly in many cases – e.g., data is leaked via email or USB stick – recent news have shown that image steganography is indeed applied to leak data out of organizational environments. Using network steganography, data leakage cannot only be realized in a stealthy but also in a constant manner, e.g. by intentionally leaking a small amount of data per hour.

## 5   How to Counter Network Steganography?

A significant problem when it comes to countermeasures is the variety of existing steganographic techniques. Not only can network steganography be applied to various network protocols and various types of digitally transmitted media (e.g. audio steams, video streams, or HTML content) but also by using an increasing number of hiding techniques. At the moment at least a few hundred hiding techniques including their combinations are known and even if digital *payload* is not taken into account, more than hundred techniques remain that transfer secret data using meta information, such as header elements or the timing of network packets [WZFH14].

Countermeasures, on the other hand, cannot address all of these available hiding techniques simultaneously due to the complexity and diversity of protocols and services and do only affect one or a few hiding techniques each. If, on the other hand, countermeasures aim on affecting many hiding methods simultaneously, it is an unsolved challenge to provide a high accuracy. The most efficient techniques applied today are the integration of traffic normalizers to prevent network steganography and the use of machine learning to detect it. Some of these techniques are integrated into data leakage protection (DLP) products available on the market.

When it comes to the application of advanced techniques as discussed in the previous section, the detection, the limitation, and the prevention, of network steganography will become even more challenging. Before countermeasures for such advanced scenarios can be built, research must come up with fundamental approaches to counter these techniques. Afterwards, products must be developed to enable a protection within organizational network environments. A recently discussed approach that is still in its infancies is to deduce so-called *patterns* (abstract descriptions) from hiding techniques and develop countermeasures for these techniques based on patterns [WZFH14]. This way, one countermeasure can influence majority of the techniques which are represented by a particular pattern.

## 6   Conclusion

The general conclusion of this work is that the network steganography methods will continue to become more and more sophisticated and harder to detect. Moreover, new and popular hidden data carriers could be exploited. Thus the network steganography threat can potentially affect every Internet user as it must be emphasized that also innocent users' traffic can be utilized by steganographers for covert communication. This will raise similar legal and ethical issues like we currently experience with botnets.

We have observed the significant increase in the application of network steganography for malicious purposes recently. Given the spectrum of available techniques that increase the quality, capabilities and stealthiness of network steganography and their dynamic evolution, it can be considered likely that further malware and malicious activities will apply network steganography "means" in an increasing manner.

A problematic aspect in this regard is the lack of effective and universal countermeasures which can be applied in practice when it comes to such sophisticated steganographic techniques. We therefore deduce a need for additional research in the field of network steganography that will lead to improved countermeasures.

**Index**

Network steganography; Information Hiding; Covert Channels; Network Security; Malware